# Superconductivity without magnetism in LiFeAs

S.V. Borisenko[1], V. B. Zabolotnyy[1], D. V. Evtushinsky[1], T. K. Kim[1], I. V. Morozov[2], A. N. Yaresko[3], A. A. Kordyuk[1], G. Behr[1], A. Vasiliev[2], R. Follath[4], B. Büchner[1]

[1]Leibniz-Institute for Solid State Research, IFW-Dresden, D-01171 Dresden, Germany
[2]Moscow State University, Moscow 119991, Russia
[3]Max-Planck-Institute for Solid State Research, D-70569 Stuttgart, Germany
[4]Helmholtz-Zentrum Berlin, BESSY, D-12489 Berlin, Germany

**The particular shape of the Fermi surface can give rise to a number of collective quantum phenomena in solids, such as density wave orderings or even superconductivity. In many new iron superconductors this shape, the "nested" Fermi surface, is indeed observed, but its role in the formation of spin-density waves or superconductivity is not clear. We have studied the electronic structure of the non-magnetic LiFeAs ($T_c$~18K) superconductor using angle-resolved photoemission spectroscopy. We find a notable absence of the Fermi surface nesting, strong renormalization of the conduction bands by a factor of three, high density of states at the Fermi level caused by a Van Hove singularity, and no evidence for either a static or fluctuating order except superconductivity with in-plane isotropic energy gaps. Our observations set a new hierarchy of the electronic properties necessary for the superconductivity in iron pnictides and, possibly, in other materials.**

The synthesis of another iron superconductor immediately attracted a considerable attention already for several reasons [1-3]. LiFeAs is one of the few superconductors which do not require additional charge carriers and has a considerable $T_c$ approaching the boiling point of hydrogen. Similar to AeFe$_2$As$_2$ (122) and LnOFeAs (1111) parent compounds, LiFeAs consists of nearly identical (Fe$_2$As$_2$)$^{2-}$ structural units and all three are isoelectronic, though the former do not superconduct. The band structure calculations unanimously yield the same shapes of the Fermi surfaces (FS), very similar densities of states and low energy electronic dispersions [4, 5, SOM] and even find in LiFeAs an energetically favorable magnetic solution which exactly corresponds to the famous stripe-like antiferromagnetic order in 122 and 1111 systems [5-7], though the experiments show rather different picture. The structural transition peculiar to 122 and 1111 families is remarkably absent in LiFeAs and is not observed under applied pressure of up to 20 GPa [8, 9]. Resistivity and susceptibility as well as µ-spin rotation experiments show no evidence for the magnetic transition [10, 11]. Only a weak magnetic background [11] and field induced magnetism in the doped compound have been detected [10]. The basic question is therefore why the isoelectronic and nearly isostructural FeAs blocks induce fundamentally different physical properties of the material? We show that this happens due to the important distinctions of the electronic structure and single out those which seem to be indispensable for the superconductivity to occur.

The knowledge gained from angle-resolved photoemission (ARPES) experiments on Fe-pnictides has often been questioned because of possible influence of the polar surface inevitably exposed by the cleavage of single crystals of 122 and 1111. LiFeAs offers a unique opportunity to overcome this difficulty since the cleavage occurs between the two layers of Li atoms resulting in equivalent and neutral counterparts.

In Fig.1a we show the FS map of LiFeAs which is the momentum distribution of the photoemission intensity integrated within the narrow energy interval around the Fermi level. Three clear features are visible on the map: a high-intensity butterfly-like shape at the $\Gamma$-point, a well defined barrel-like FS also centred at $\Gamma$, and a double-walled structure with somewhat obscure contours around M-point. Momentum-energy cuts along the selected directions passing near M and $\Gamma$ points and marked in panel a) by red arrows are presented in Fig.1b and Fig.1c. While both dispersive features in Fig.1b as well as the left-most one in Fig.1c clearly cross the Fermi level thus supporting a double-walled electron-like FS around M and large hole-like FS around $\Gamma$ respectively, the other two features from Fig.1c only come close to the Fermi level without clear crossing, at least for the given cut through the momentum space. We have found that for certain excitation energies these features do cross the Fermi level resulting in very small hole-like FSs thus completing the analogy with the FS topolgy of 122 and 1111 systems: there are in total five FSs supported by five bands. The sketch in Fig.1a schematically shows all FS features centered in one point to facilitate the comparison of their sizes. The striking peculiarity of this FS shape is the absolute absence of the $(\pi, \pi)$-nesting peculiar to other pnictides.

According to the existent band structure calculations, including our own study (see SOM), the FS of LiFeAs also consists of three hole-like FSs around the $\Gamma$-point and two electron-like ones at the corner of the Brillouin zone. Although the three-dimensional character of the FS is less pronounced in comparison with 122 and 1111 systems, the band responsible for the smallest FS at $\Gamma$ crosses the Fermi level along $\Gamma Z$ direction and the second largest FS near $\Gamma$ and both electron-like FSs near M exhibit noticeable dependence on $k_z$. In Fig.2 we compare our experimental data with the calculations. Three FS maps shown in Fig.2a-c are recorded at different photon energies which correspond to different $k_z$ values and thus should reflect the variations of the FS with $k_z$. Indeed, being well defined and having the same size in all maps, the largest $\Gamma$-barrel is nearly ideally two-dimensional in close agreement with the calculations. At the same time the degree of three-dimensionality of the two small $\Gamma$-centred FSs is overestimated in the calculations. The distribution of the spectral weight around M-point is photon energy dependent and takes the shape of either crossed ellipses or rounded squares or displays the signatures of both (see also Fig.1a), more or less exactly following the theoretical predictions.

To understand in more details the butterfly-like intensity spot at $\Gamma$, which universally appears in all maps and looks different only in replicated $\Gamma$-points at non-zero momentum, we performed the polarisation dependent measurements. The strong influence of the matrix elements on the photoemission intensity seen in Fig.2d is expected and fully explainable in terms of symmetry composition of the states forming two small FSs at $\Gamma$-point. According to calculations [SOM] these states are formed by $d_{xz}$ ($d_{yz}$) orbitals and are therefore enhanced (suppressed) when probed in $\pi$ ($\sigma$) polarisation geometries [12]. Note that the behavior of the outer $\Gamma$-barrel is also easy to understand, as it is formed by the states having $d_{x^2-y^2}$ symmetry and thus also sensitive to such a variation of the geometry of the experiment. From Fig.2d (see also Fig.1c) it also follows that the tops of both bands practically touch the Fermi level with the top of the larger one having an extended flat shape. This results in an anomalous enhancement of the density of states at the Fermi level.

Knowing the shape and topology of the FS in details we were able to estimate the number of charge carriers in the system. Neglecting the contribution of the smallest $\Gamma$-FS it turns out that two electron-like pockets compensate two hole-like $\Gamma$-FSs and the total occupation of the corresponding four bands roughly equals four electrons. More rigorous estimates which take into account $k_Z$ dispersion of the electron pockets also give the stoichiometric solution within the error bars. The measurements have been carried out on three different single crystals and yielded reproducible results thus indicating the absence of noticeable Li content variations (see SOM).

In Fig.2e we compare directly the calculated band structure and experimental data along the high-symmetry directions. We made a tight-binding fit to the best defined experimental band which supports the large $\Gamma$-barrel and found a remarkable qualitative agreement with the $d_{x^2-y^2}$ originated band from the LDA calculations. The ratio of the bandwidths turned out to be equal to 3.1 (1 eV vs. 0.326 eV). We then applied this renormalization factor to bring the experimental data and the calculated bands to the same energy scale. Overall agreement is very reasonable, taking into account that we have identified all bands in the vicinity of the Fermi level and associated them with the corresponding features in the ARPES spectra. Note that the electron-like pocket at M formed by another band is satisfactorily reproduced as well. As seen from the figure and confirmed by the calculations, a selective shift of the $d_{x^2-y^2}$ band (red curve in Fig.2e) by 150 meV with respect to the others perfectly reproduces the shape of the experimental Fermi surface and dispersion over the whole bandwidth.

This agreement together with the extremely low temperatures at which we carried out our experiments clearly speaks in favor of a non-magnetic ground state realized in LiFeAs. This is supported by the fact that having explored the large portions of the k-space at different experimental conditions, we have not found any typical spectroscopic signatures of commensurate or incommesurate ordering appearing in a form of replica due to Fermi surface reconstruction or suppressed spectral weight due to gap, etc. [13, 14]. What is not captured by the LDA approach, which predicts magnetism in LiFeAs [5, 6], are the actual absence of nesting and renormalization by the factor of ~3. The energy gain from the opening of the gaps at or near the Fermi level due to the FS reconstruction is obviously more significant in a system with 3 times narrower bandwidth. That is probably why the SDW order does not disappear immediately upon doping in 122 and 1111 families of Fe-pnictides. In LiFeAs, where the $(\pi, \pi)$-nesting is absent, static magnetism disappears completely. Our results thus strongly imply the decisive role of nesting in the formation of SDW.

The apparent absence of the magnetism in LiFeAs seems to be not crucial for the superconductivity. We have clearly observed the opening of the superconducting gap in all $k_F$ points with the onset at the nominal $T_c$ of ~18 K. A typical example is presented in Fig.3a where the crossing of the FS at point A (see Fig.3d) is shown for two temperatures, above and below the transition. Upon entering the superconducting state, the usual BCS-like bending back of the dispersive feature is apparently seen in the lower panel and the experimental dispersion, determined as a set of maxima of the momentum distribution curves, exhibits peculiar transformation from the straight line to the "S"-shaped curve in the immediate vicinity of the Fermi level (not shown). This is accompanied by the depletion of the spectral weight at the Fermi level with concomitant shift of the leading edge midpoint of the $k_F$-EDC as seen in Fig.3b and in zoomed version in Fig.3c. In passing, there are well defined quasiparticle excitations, both above and below Tc (Fig.3b) as well as characteristic "kinks" in the dispersion at 18 meV and 30 meV. We conclude the presentation of the experimental material by plotting the kinetic energies of the leading edge midpoints as a function of momentum in Fig.3d. The typical size of the superconducting gap of the hole-like FSs is estimated to be ~ 1.5 - 2.5 meV, while for the electron-like FSs we observed slightly larger values of ~ 2 – 3.5 meV implying a multiband superconductivity in LiFeAs. In both cases, as follows from Fig.3d, the anisotropy of the gap is not dramatic, although taking into account the relatively small absolute values of the gaps, even small variations may be of interest. More rigorous estimates based on accurate fitting including the resolution contribution result in the superconducting gap of 3.2 meV on the electron-like pockets.

Disregarding for a moment any relation to other families of Fe-pnictides and thus possible presence of the so far not detected spin fluctuations capable of binding electrons in pairs [15], one may ask why a stoichiometric (or very close to that), nonmagnetic compound with a plane crystal structure is a multiband superconductor with considerable critical temperature and weakly in-plane anisotropic order parameter? The only remarkable property of the electronic structure of LiFeAs which makes it so special and intimately connects it to other superconductors like $NbSe_2$, A15 compounds, $MgB_2$ and the cuprates is the proximity of the Van Hove singularity to the Fermi level. Coming back to other Fe-pnictides, one can easily reconcile an accidental enhancement of the density of states at the Fermi level in LiFeAs with the similar enhancement achieved by doping in hole- or electron-doped 122 and 1111 systems (Fig.4). It is important to note that merely the extremum of a paraboloid-like band touching the Fermi level will not result in DOS boost at the Fermi level in a two-dimensional material. It has to be either a saddle point or an extended flat region of the dispersion, as is the case in the electronic structures sketched in Fig. 4.

In fact, it is the presence of the Van Hove singularity in LiFeAs which is most likely responsible for the strong renormalization of the conduction band observed in the present work. The enhanced quasiparticle scattering is indeed expected because of the much less restricted phase space for scattering provided by the states near the Fermi level [16]. We do find unusually steep increase of the imaginary part of the self-energy which naturally agrees with the considerable real part responsible for the renormalization. A remarkable analogy with dichalcogenides is clearly in favour of this conclusion. The experimental bandwidth of 2H-$TaSe_2$ remains practically unrenormalized compared to the LDA results [13], whereas in a very similar 2H-$NbSe_2$ the renormalization is of the factor of two [17]. One should point out that going from Nb (4d) to Ta (5d) the localization of the valence d-shell is not changed significantly (less than 4% in atoms) and the considerable renormalization cannot be attributed to the larger U. Thus, the only principal difference between the electronic structures of these compounds is the energetic location of the saddle point between Γ and K points (Fig. 4). In the former it is above the Fermi level while in the latter it is just below.

The example of LiFeAs suggests that the intriguing presence of the density waves or their fluctuations in many families of the superconductors may turn out to be rather "technical": they neighbour the superconductivity only because, purely from the topological considerations, the nesting is very probable to occur when tuning the system from one Van Hove instability to another by doping in multiband superconductors (see Fig.4). On the other hand, the non-perfect nesting corresponding to the fading density wave order may result in still larger density of states at the Fermi level thus promoting the superconductivity [18]. However, in view of our present results, the Van Hove singularity close to the Fermi level seems to be a necessary condition for the onset of superconductivity, and not only in the pnictides.

ACKNOWLEDGEMENTS
We acknowledge fruitful discussions with Ernst Bauer, Sergey Bud'ko, Ralf Claessen, Ilya Eremin, Hideo Hosono, Arno Kampf, Dirk van der Marel, Joel Mesot and Werner Weber. The project was supported, in part, by the DFG under Grant No. KN393/4 and BO 1912/2-1. I.V. Morozov also acknowledges support from the Ministry of Science and Education of the Russian Federation under State Contract P-279. We are grateful to S. Graser for sharing with us his results of the band structure calculations.

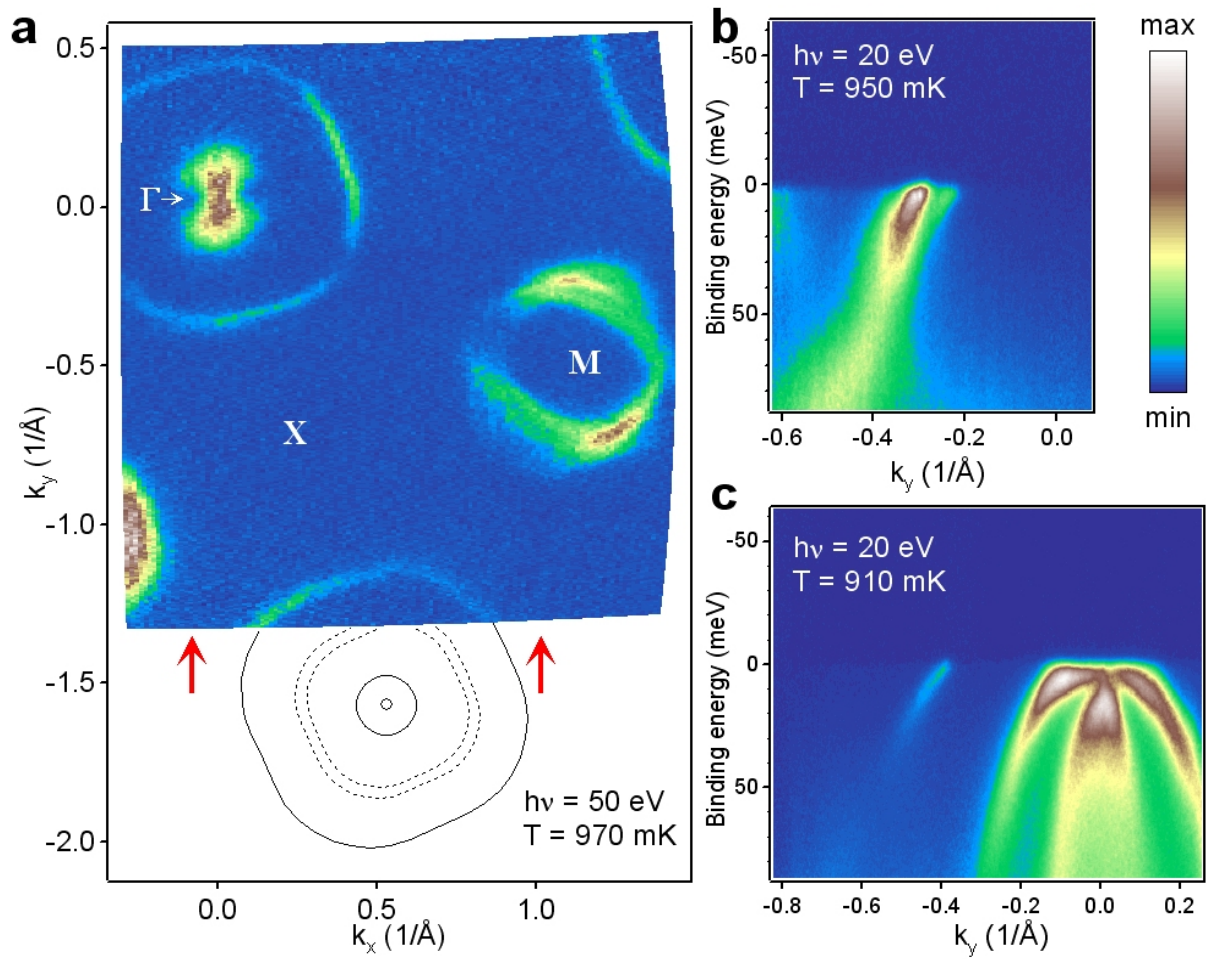

**Figure1.** Fermi surface shape and topology. **a)** Momentum distribution map of the photoemission intensity integrated within 5 meV around the Fermi level. Solid lines represent Γ - centered Fermi contours, dashed lines – most pronounced M-centered Fermi contours. **b,c)** Momentum-energy cuts along the vertical directions marked by the red arrows in panel a).

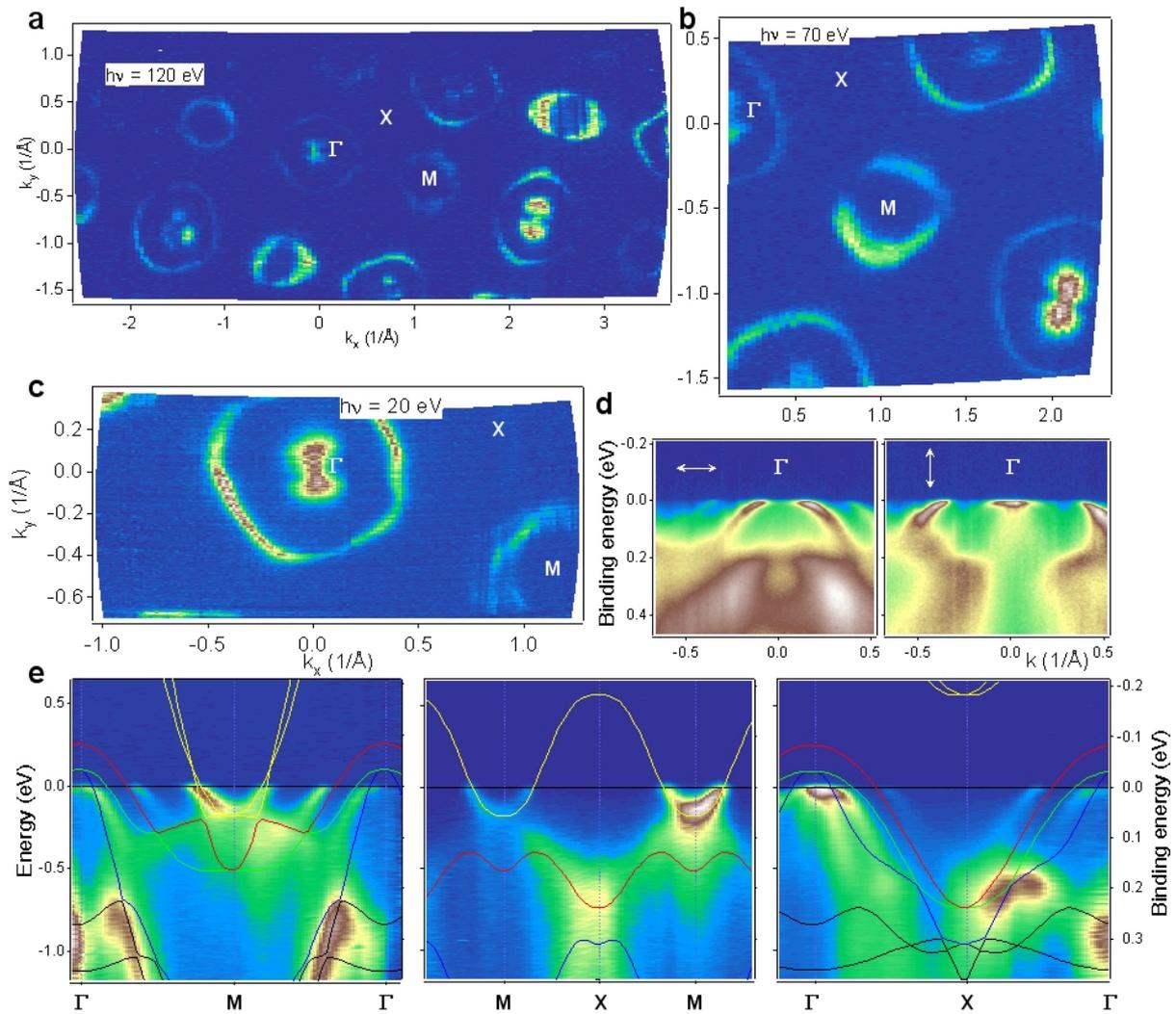

**Figure 2.** Comparison with the LDA calculations. **a, b, c)** Fermi surface maps taken at different excitation energies at T~1K. **d)** Momentum-energy cut through the Γ-point measured with the σ (left) and π (right) polarisations. Photon energy is 67 eV. **e)** High-symmetry momentum-energy cuts together with the results of the band structure calculations. Blue, green and red curves are the bands supporting the Γ-centred FSs. Yellow curves – the bands responsible for the electron-like FSs. Black curves – the bands which do not cross the Fermi level. Photon energy is 120 eV for the middle panel and 70 eV for the left and right panels.

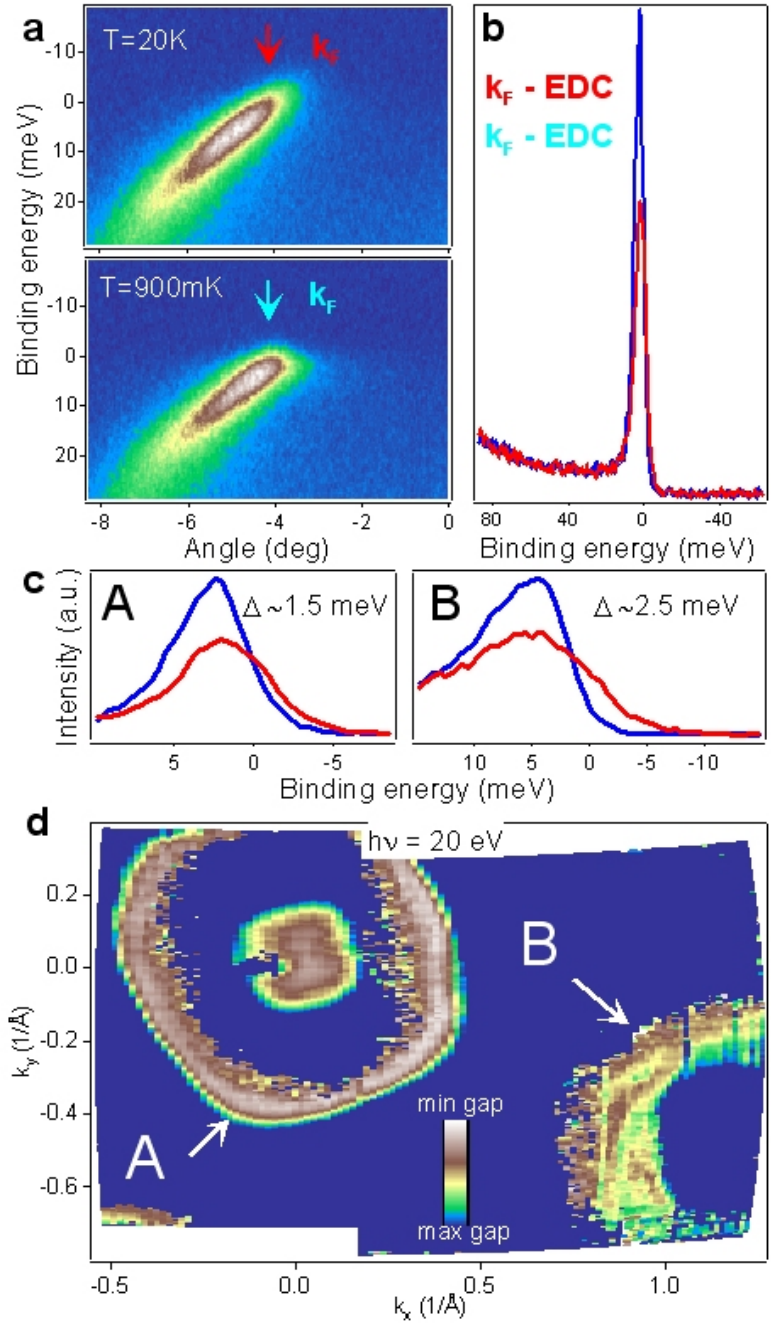

**Figure 3.** Superconductivity in LiFeAs. **a)** Fermi surface crossings at the point A (see panel d) taken above and below $T_c$ (~18K). **b)** Energy distribution curves from panels a) corresponding to the $k_F$. **c)** Energy distribution curves corresponding to Fermi momenta A and B measured above (20 K) and below (900 mK) $T_c$. **d)** Kinetic energies of the leading edge midpoints of all energy distribution curves from the shown momentum space region in the false color scale. The local maxima of this distribution correspond to $k_F$ points also seen in Fig.2c and give a good measure of the superconducting gap.

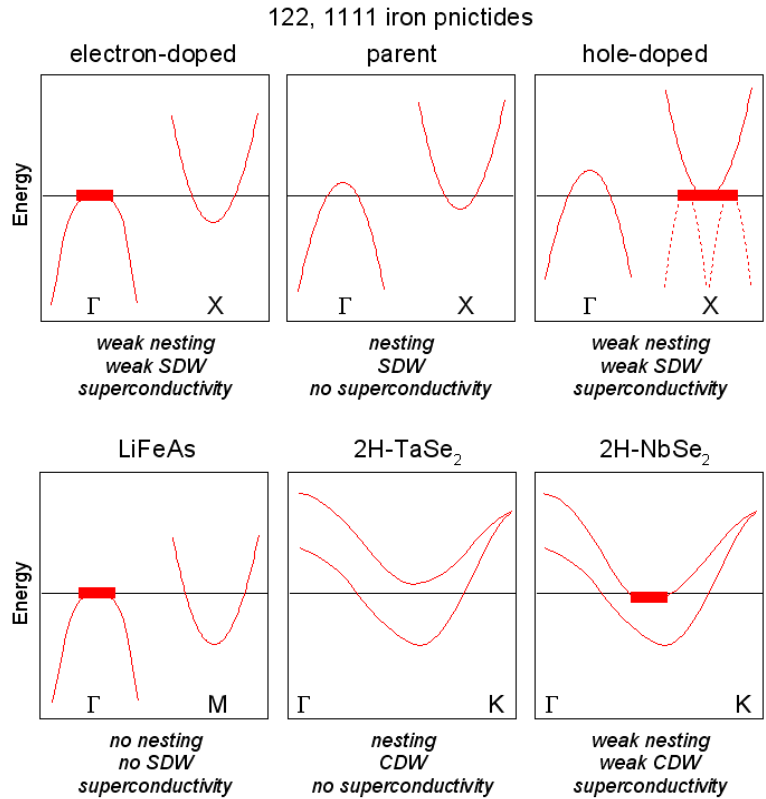

**Figure 4.** Van Hove singularities in Fe-pnictides and chalcogenides. Schematic illustration of the main features in the electronic structure of the 122 and 1111 pnictides, LiFeAs, 2H-TaSe$_2$ and 2H-NbSe$_2$. Red curves represent the band dispersions along the high-symmetry directions. Thick red lines represent Van Hove sigularities, i.e. the flat extrema of the corresponding bands or the saddle point.

## SUPPORTING ONLINE MATERIAL

**Single crystals**

Three large ( ~ 3x2x0.5mm) single crystals of LiFeAs have been used in the current study, two Li11(SF) and one Li11(Sn). All operations on preparation of the large LiFeAs single crystals have been carried out in a dry box in Ar atmosphere. To grow the single crystal of LiFeAs from the melted tin (sample Li11(Sn)) the reagents in molar ratio Li:Fe:As:Sn=1.15:1:1:24 were mixed in alumina crucible which was inserted into Nb container sealed under 1.5 atm of argon gas. The Nb container was then sealed in an evacuated quartz ampoule and heated to 1163 K, and after that it was slowly cooled down to 853 K. At this temperature, the liquid tin was removed by decantation and than by centrifugation at high temperature. To grow the LiFeAs single crystals by self-flux method (Li11(SF)) the reagents in molar ratio Li:Fe:As = 3:2:3 were inserted to the same package heated up to 1363 K, kept at this temperature for 5 hours and cooled at a rate 4.5 K/h down to 873 K and than the furnace was switched on. The plate-like single crystals were separated from the flux mechanically. Phase identification was performed by means of X-Ray powder diffraction analysis of the polycrystalline samples, prepared from the single crystals by grinding them in a dry box. The results of tetragonal P4/nmm unit cell refinement ($a$=3.7701(15), $c$=6.3512(25) Å, V=90.27(8) Å$^3$ for Li11(SF) and $a$=3.7680(24), $c$=6.339(4)Å, V=90.00(13) Å$^3$ for Li11(Sn)) are in good agreement with the data available in the literature [2, 3]. The molar ratio of Fe:As close to 1:1, as well as the existence of about 0.5 mol. % of tin in the Li11(Sn) crystals have been identified from the EDX data.

**ARPES**

Photoemission experiments have been carried out using the synchrotron radiation from the BESSY storage ring. The end-station "1-cubed ARPES" is equipped with the He3 cryostat which allows to collect the angle-resolved spectra at temperatures below 1K. The overall energy resolution ranged from ~2.5 meV at h$\nu$ = 15 eV to ~ 6 meV at h$\nu$ =120 eV. All single crystals have been cleaved in UHV exposing the mirror-like surfaces.

**Band structure calculations**

Band structure of LiFeAs has been calculated for the experimental crystal structure [3] using the LMTO method in the atomic sphere approximation. "Fat" bands and FS cross-section obtained from the spin-restricted LDA calculation are shown in Figure 1 (SOM). The largest (red) hole-like Fermi surface centered at the $\Gamma$ point is formed by the band originating from Fe $x^2$-$y^2$ states. Since the $x^2$-$y^2$ states are even with respect to reflection in the zx mirror plane, the corresponding band can be observed in the ARPES experiment only when probed by photons with the polarisation vector lying in the mirror plane ($\pi$ polarisation). The other two hole-like FSs are formed by Fe xz,yz-derived bands, which are degenerate at the $\Gamma$ point. As one moves away from $\Gamma$ along the $\Gamma$-X line, the band responsible for the innermost (blue) FS acquires increasing contribution of the even Fe 3$z^2$-1 states (Figure 1 SOM a). In contrast, another (green) band shows appreciable admixture of the Fe xy states, which are odd with respect to the xz mirror plane. Thus, along the $\Gamma$-X line, or, more generally, in the $k_x$-$k_z$ plane, the former of these bands is formed by the even Fe xz states and is observed together the $x^2-y^2$ band in the π polarization. The latter band is dominated by the odd Fe yz states and can be probed only by photons with the polarisation vector perpendicular to the mirror plane (σ polarisation).

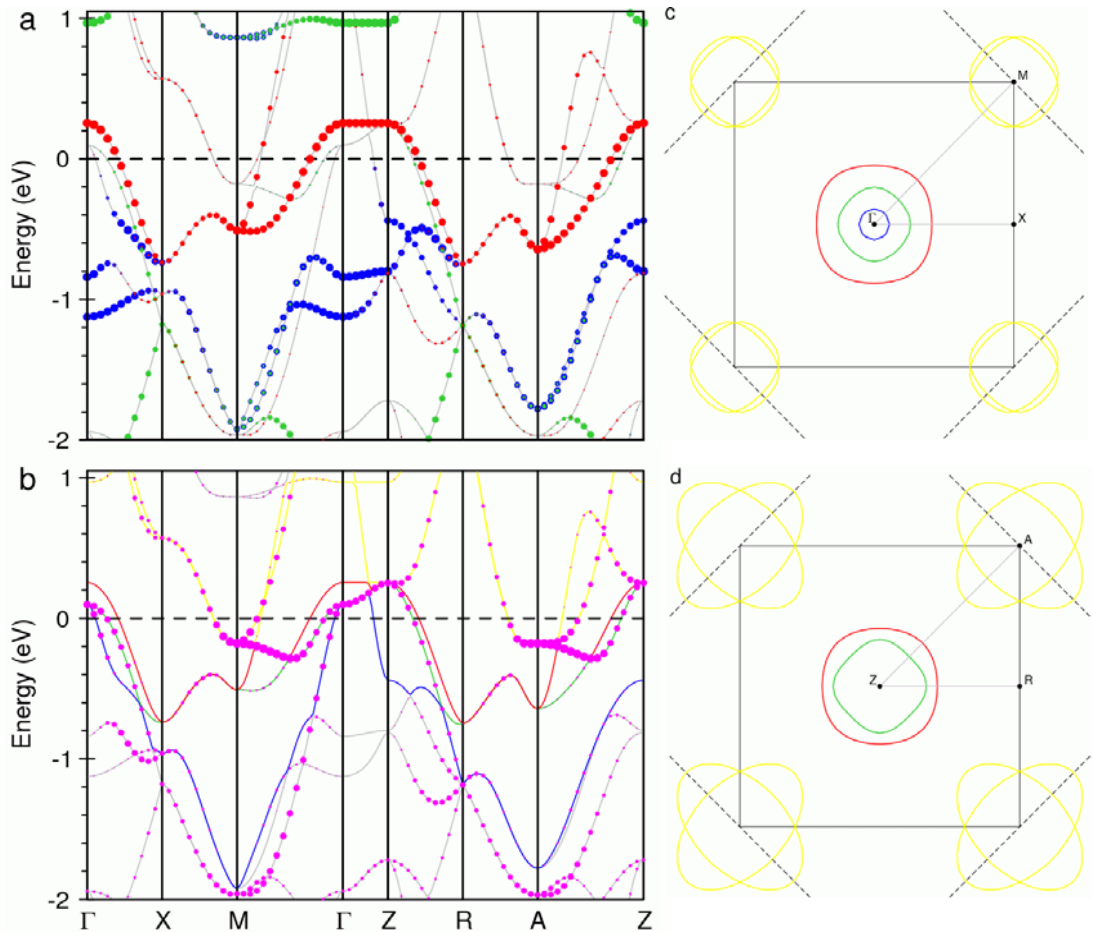

**Figure 1 SOM.** a,b) "Fat" bands obtained from the spin-restricted LDA calculation for LiFeAs. The size of colored circles in the plots is proportional to the partial weight of a particular state in the Bloch wave function. In panel a) the partial weights of the Fe d $3z^2-1$, xy, and $x^2-y^2$ states are plotted by blue, green, and red circles, respectively. The sum of the contributions of the Fe d xz and yz states is plotted in panel (b) by magenta circles. The bands forming the hole-like FS sheets around the Γ point are shown by blue, green, and red lines, whereas the bands responsible for the M-centered electron-like FS are plotted by yellow lines. c,d) Calculated cross-sections of the Fermi surface by the ΓXM (c) and ZRA (d) planes. The cross-sections originating from various hole-like and electron-like bands are plotted with the same color as in panel (b).